\documentclass[12pt]{article}

\usepackage{algorithm}				
\usepackage[noend]{algpseudocode}	
\usepackage{amsmath}				
\usepackage{amssymb}				
\usepackage{amsthm}				
\usepackage{caption}
\usepackage{subcaption}
\usepackage{color}
\usepackage[margin=1in]{geometry}
\usepackage[hidelinks]{hyperref}
\usepackage{titling}

\usepackage{tikz}
\usetikzlibrary{matrix}

\usepackage{pifont}				
\newcommand{\cmark}{\ding{51}}
\newcommand{\xmark}{\ding{55}}

\newcommand*\circled[1]{\tikz[baseline=(char.base)]{\node[shape=circle,draw,inner sep=0.5pt] (char) {#1};}}

\usepackage{stmaryrd}			

\theoremstyle{plain}				
\newtheorem{theorem}{Theorem}
\newtheorem{lemma}[theorem]{Lemma}

\newtheorem{corollary}[theorem]{Corollary}

\theoremstyle{definition}			
\newtheorem{definition}[theorem]{Definition}

\theoremstyle{remark}			
\newtheorem*{remark}{Remark}

\newcommand{\TDFA}{\textsf{2DFA-4W}}

\newcommand{\TDFATW}{\textsf{2DFA-3W}}

\newcommand{\TDFATWOW}{\textsf{2DFA-2W}}

\newcommand{\TNFA}{\textsf{2NFA-4W}}

\newcommand{\TNFATW}{\textsf{2NFA-3W}}

\newcommand{\TNFATWOW}{\textsf{2NFA-2W}}

\thanksmarkseries{alph}

\title{Concatenation Operations and Restricted Variants of Two-Dimensional Automata}
\author{Taylor J. Smith \thanks{School of Computing, Queen's University, Kingston, Ontario, Canada. Email: \texttt{\{tsmith,ksalomaa\}@cs.queensu.ca}.} \and Kai Salomaa \thanksmark{1}}
\date{\today}

\begin{document}


\maketitle

\begin{abstract}
A two-dimensional automaton 
operates on arrays of symbols. While a standard (four-way) two-dimensional automaton can move its input head in four directions, restricted two-dimensional automata are only permitted to move their input heads in three or two directions; these models are called three-way and two-way two-dimensional automata, respectively.

In two dimensions, we may extend the notion of concatenation in multiple ways, depending on the words to be concatenated. We may row-concatenate (resp., column-concatenate) a pair of two-dimensional words when they have the same number of columns (resp., rows). In addition, the diagonal concatenation operation combines two words at their lower-right and upper-left corners, and is not dimension-dependent.

In this paper, we investigate closure properties of restricted models of two-dimensional automata under various concatenation operations. We give non-closure results for two-way two-dimensional automata under row and column concatenation in both the deterministic and nondeterministic cases. We further give positive closure results for the same concatenation operations on unary nondeterministic two-way two-dimensional automata. Finally, we study closure properties of diagonal concatenation on both two- and three-way two-dimensional automata.

\medskip


\medskip

\noindent\textit{MSC2020 classes:} 68Q45 (primary); 68R15 (secondary).
\end{abstract}


\section{Introduction}\label{sec:introduction}

The two-dimensional automaton model, introduced by Blum and Hewitt \cite{BlumHewitt19672DAutomata}, is a generalization of the well-known one-dimensional (string) automaton model. A two-dimensional automaton takes as input an array or matrix of symbols from some alphabet $\Sigma$, and the input head of the automaton moves in four directions: upward, downward, leftward, and rightward.

If we restrict the input head movement of a two-dimensional automaton, then we obtain a variant of the model that is weaker in terms of recognition power, but also easier to reason about. If we prevent the input head from moving upward, then we obtain a three-way two-dimensional automaton. If we prevent both upward and leftward moves, then we obtain a two-way two-dimensional automaton. The three-way two-dimensional automaton model was introduced by Rosenfeld \cite{Rosenfeld1979PictureLanguages}. The two-way two-dimensional automaton model was introduced by Anselmo et al.\ \cite{Anselmo2005NewOperations2DLanguages} and formalized by Dong and Jin \cite{Dong2012TwoWay2DAutomata}.

We can generalize many language operations from one dimension to two dimensions, and most of these operations have been studied in the past; for a review of previous work, see the surveys by Inoue and Takanami \cite{Inoue19912DAutomataSurvey} or by the first author \cite{Smith2019TwoDimensionalAutomata}. In this paper, we focus on the language operation of concatenation. In two dimensions, we may concatenate words either by joining rows or by joining columns. In either case, the relevant dimension of the words being concatenated must be equal (e.g., two words being concatenated column-wise must have the same number of rows).

Four-way two-dimensional automata are not closed under either row or column concatenation \cite{Inoue1978Note2DAutomata}. Three-way two-dimensional automata are not closed under column concatenation, but nondeterministic three-way two-dimensional automata are closed under row concatenation \cite{InoueTakanami1979ClosureProperties2DTuring}. A selection of known closure results is summarized in Table~\ref{tab:2Dclosure}.

In this paper, we investigate the closure of restricted two-dimensional automaton models under various concatenation operations. We give the first closure results for concatenation of languages recognized by two-way two-dimensional automata, showing that the model is not closed under row or column concatenation in the general alphabet case, while it is closed under both operations in the unary nondeterministic case. 
After defining a third method of concatenation known as ``diagonal concatenation", we show that nondeterministic two-way two-dimensional automata are closed under this operation, while this closure is lost in the deterministic case. Finally, we prove that deterministic three-way two-dimensional automata are also not closed under diagonal concatenation.

\begin{table}[t]
\centering
\begin{tabular}{l | c c c c c c}
					& \TDFA			& \TNFA			& \TDFATW		& \TNFATW 		& \TDFATWOW		& \TNFATWOW \\
\hline
Row ($\ominus$)	& \xmark 			& \xmark 			& \xmark			& \cmark 			& \circled{\xmark}	& \circled{\xmark} / \circled{\cmark}$^{\, \dagger}$ \\
Column ($\obar$)		& \xmark 			& \xmark 			& \xmark			& \xmark 			& \circled{\xmark}	& \circled{\xmark} / \circled{\cmark}$^{\, \dagger}$ \\
Diagonal ($\oslash$)		& \textbf{?}		& \textbf{?}		& \circled{\xmark}	& \textbf{?}		& \circled{\xmark}	& \circled{\cmark}
\end{tabular}
\caption{Closure results for concatenation on two-dimensional automaton models. Closure is denoted by \cmark\ and nonclosure is denoted by \xmark. New closure results presented in this paper are circled. Closure results marked with a $^{\dagger}$ apply in the unary case.}
\label{tab:2Dclosure}
\end{table}


\section{Preliminaries}\label{sec:preliminaries}

A two-dimensional word consists of a finite array, or rectangle, of cells each labelled by a symbol from a finite alphabet $\Sigma$. 
Precisely speaking, for $m, n \geq 1$, an $m \times n$ two-dimensional word is a map from $\{1, \dots, m\} \times \{1, \dots, n\}$ to $\Sigma$. 
When a two-dimensional word is written on an input tape, 
the cells around the two-dimensional word are labelled with a special boundary marker $\# \not\in \Sigma$; 
more generally, we may consider all cells outside of the bounds of an input word to contain boundary markers (see Kari and Salo~\cite{KariSalo2011PictureWalkingAutomataSurvey}, particularly Sections~2 and 4).

A two-dimensional automaton has a finite state control that is capable of moving its input head in four directions within an input word: up, down, left, and right. We denote these directions by $U$, $D$, $L$, and $R$, respectively.

\begin{definition}[Two-dimensional automaton]\label{def:2DFA}
A two-dimensional 
automaton is a tuple \allowbreak $(Q, \Sigma, \delta, q_{0}, q_{\rm accept})$, where $Q$ is a finite set of states, $\Sigma$ is the input alphabet (with $\# \not\in \Sigma$ acting as a boundary symbol), $\delta: (Q \setminus \{q_{\rm accept}\}) \times (\Sigma \cup \{\#\}) \to Q \times \{U, D, L, R\}$ is the partial transition function, and $q_{0}, q_{\rm accept} \in Q$ are the initial and accepting states, respectively.
\end{definition}

The computation of a two-dimensional automaton begins in the top-left corner 
(i.e., at cell $(1,1)$) 
of its input word in the initial state $q_{0}$, and the automaton halts and accepts when it reaches the accepting state $q_{\rm accept}$.

We can modify the deterministic model given in Definition~\ref{def:2DFA} to be nondeterministic by changing the transition function to map to $2^{Q \times \{U, D, L, R\}}$ instead of $Q \times \{U, D, L, R\}$. We denote the deterministic and nondeterministic two-dimensional automaton models by \TDFA\ and \TNFA, respectively, where \textsf{4W} indicates that the automaton has four directions of movement.

By restricting the movement of the input head to move in fewer than four directions, we obtain the aforementioned restricted variants of the two-dimensional automaton model. If we prohibit upward movements, then we get a three-way two-dimensional automaton. If we prohibit both upward and leftward movements, then we get a two-way two-dimensional automaton.

\begin{definition}[Three-way/two-way two-dimensional automaton]
A three-way (resp., two-way) two-dimensional automaton is a tuple $(Q, \Sigma, \delta, q_{0}, q_{\rm accept})$ as in Definition~\ref{def:2DFA}, where the transition function $\delta$ is restricted to use only the directions $\{D, L, R\}$ (resp., the directions $\{D, R\}$).
\end{definition}

We denote three-way two-dimensional automata by \TDFATW\ / \TNFATW, and we denote two-way two-dimensional automata by \TDFATWOW\ / \TNFATWOW. Note that three-way two-dimensional automata are unable to move their input head back into a word upon leaving the bottom edge of the word, while two-way two-dimensional automata are unable to do the same upon leaving either the bottom or right edge of the word. Thus, if a three-way (resp., two-way) two-dimensional automaton makes a downward (resp., downward or rightward) move and reads a boundary symbol, it can only read boundary symbols for the remainder of its computation.


\section{Row and Column Concatenation}\label{sec:hvconcatenation}

In two dimensions, we may consider the notions of row and column concatenation. The row (resp., column) concatenation of two-dimensional words $w$ and $v$, denoted $w \ominus v$ (resp., $w \obar v$), is the word produced by adjoining the last row (resp., column) of $w$ to the first row (resp., column) of $v$. 
If $w$ and $v$ are of dimension $m \times n$ and $m' \times n'$ respectively, then the row and column concatenations of these words are defined only when $n = n'$ or $m = m'$, respectively.

We may similarly define the row or column concatenation of two languages $A$ and $B$ as $A \circ B = \{a \circ b \mid a \in A \text{ and } b \in B\}$, where $\circ \in \{\ominus, \obar\}$.

\begin{figure}[t]
\[\arraycolsep=1.4pt\def\arraystretch{0.8}
w \ominus v = 
\begin{array}{ccccc}
\#		& \#		& 		& \#		& \# \\
\#		& w_{1,1}	& \cdots	& w_{1,n}	& \# \\
		& \vdots	& \ddots	& \vdots	& \\
\#		& w_{m,1}	& \cdots	& w_{m,n}	& \# \\
\#		& v_{1,1}	& \cdots	& v_{1,n}	& \# \\
		& \vdots	& \ddots	& \vdots	& \\
\#		& v_{m',1}	& \cdots	& v_{m',n}	& \# \\
\#		& \#		& 		& \#		& \#
\end{array}
\hspace{1cm}%
w \obar v = 
\begin{array}{cccccccc}
\#		& \#		& 		& \#			& \#		& 		& \# 		& \# \\
\#		& w_{1,1}	& \cdots	& w_{1,n}		& v_{1,1}	& \cdots	& v_{1,n'}	& \# \\
		& \vdots	& \ddots	& \vdots		& \vdots	& \ddots	& \vdots	& \\
\#		& w_{m,1}	& \cdots	& w_{m,n}		& v_{m,1}	& \cdots	& v_{m,n'}	& \# \\
\#		& \#		& 		& \#			& \#		& 		& \#		& \#
\end{array} 
\]
\caption{Row and column concatenations of two-dimensional words}
\label{fig:2Dconcatenation}
\end{figure}

Figure~\ref{fig:2Dconcatenation} illustrates the row and column concatenations of two words.


\subsection{Two-Way Two-Dimensional Automata}

As we noted in the introduction, basic closure results about concatenation are known for both four-way and three-way two-dimensional automata, and the only positive closure result applies to row concatenation over nondeterministic three-way two-dimensional automata. 
Unfortunately, for the two-way two-dimensional automaton model, 
we do not have closure for either row or column concatenation in the general alphabet case. Here, we 
state the result for the row concatenation operation on nondeterministic two-way two-dimensional automata.

\begin{theorem}\label{thm:2Wrowconcatenation}
Nondeterministic two-way two-dimensional automata over a general alphabet are not closed under row concatenation.

\begin{proof}
Define a language $L$ as the set of two-dimensional words over the alphabet $\Sigma = \{\texttt{0}, \texttt{1}\}$ where the first row of each word consists only of the symbol \texttt{0}. The language $L$ can be recognized by a two-way two-dimensional automaton whose input head scans the first row of the input word.

Suppose there exists a nondeterministic two-way two-dimensional automaton $\mathcal{A}$ that recognizes the language $L \ominus L$. Then $\mathcal{A}$ accepts an input word $w$ of dimension $2 \times 2$ consisting entirely of the symbol \texttt{0}. Clearly, $w \in L \ominus L$. However, the accepting computation of $\mathcal{A}$ on $w$ cannot visit all four symbols in the word, and thus $\mathcal{A}$ necessarily also accepts another $2 \times 2$ input word $v$ that contains one occurrence of the symbol \texttt{1} in a cell not visited during the accepting computation. Since $v \not\in L \ominus L$, this is a contradiction.
\end{proof}
\end{theorem}

The preceding theorem 
can easily be adapted to work 
in the deterministic case, and 
we can similarly prove non-closure for column concatenation over two-way two-dimensional automata.

Altogether, the previous results show that general-alphabet languages recognized by a deterministic two-way two-dimensional automaton may be concatenated either row-wise or column-wise to produce a language not recognized even by a nondeterministic two-way two-dimensional automaton.


\subsection{Unary Two-Way Two-Dimensional Automata}

As a consequence of the closure of nondeterministic three-way two-dimensional automata under row concatenation, 
we also know that unary nondeterministic three-way two-dimensional automata are closed under this operation. Aside from this fact, not much is known about closure of concatenation for unary two-dimensional automaton models. In this section, we obtain new closure results for row and column concatenation for unary nondeterministic two-way two-dimensional automata.

\begin{remark}
Anselmo et al.\ \cite{Anselmo2005NewOperations2DLanguages} previously studied properties of the two-way two-dimensional automaton model over a unary alphabet; however, their model differs from the one considered in this paper. See Section~\ref{subsec:dconcatenationtwoway} for more details.
\end{remark}

Before we proceed, we require one further definition. 
We say that an automaton is ``immediately BR-accepting", or ``IBR-accepting", if, upon reading a boundary marker on the bottom or right border of the word, the automaton immediately halts and accepts if $q_{\text{accept}}$ is reachable from its current state.

\begin{lemma}
Given a two-way two-dimensional automaton $\mathcal{M}$, there exists an equivalent IBR-accepting two-way two-dimensional automaton $\mathcal{M}'$.

\begin{proof}
If $\mathcal{M}$ reads a boundary marker at the bottom or right border of its input word, then the input head of $\mathcal{M}$ can only read boundary markers for the remainder of its computation. After reading a boundary marker in state $q_{i}$, say, we can decide whether $q_{\text{accept}}$ is reachable from $q_{i}$ 
via some sequence of transitions on an arbitrary number of boundary markers. Thus, we may take $\mathcal{M}'$ to be the same as $\mathcal{M}$ apart from its transition upon reading the first boundary marker, which we modify to transition to $q_{\text{accept}}$ in the positive case or leave undefined in the negative case.
\end{proof}
\end{lemma}

Using the notion of an IBR-accepting automaton, we obtain the main result of the section.

\begin{theorem}
Nondeterministic two-way two-dimensional automata over a unary alphabet are closed under row concatenation.

\begin{proof}
Let $\mathcal{A}$ and $\mathcal{B}$ be unary nondeterministic two-way two-dimensional automata recognizing languages $A$ and $B$, respectively. Assume both $\mathcal{A}$ and $\mathcal{B}$ are IBR-accepting, and let the accepting computations of $\mathcal{A}$ and $\mathcal{B}$ be denoted by $C_{\mathcal{A}}$ and $C_{\mathcal{B}}$, respectively. We construct another unary nondeterministic two-way two-dimensional automaton $\mathcal{M}$ to recognize the language $A \ominus B$. 
The automaton $\mathcal{M}$ first makes a nondeterministic choice of which ``types" of computation correspond to $C_{\mathcal{A}}$ and $C_{\mathcal{B}}$, and then interleaves both computations. The cases for each ``type" of computation are as follows:
\begin{enumerate}
\item\label{enum:bottomright} $C_{\mathcal{A}}$ accepts at the bottom border and $C_{\mathcal{B}}$ accepts at the right border;
\item\label{enum:rightbottom} $C_{\mathcal{A}}$ accepts at the right border and $C_{\mathcal{B}}$ accepts at the bottom border;
\item 
\begin{enumerate}
\item\label{enum:bottombefore} $C_{\mathcal{A}}$ accepts at the bottom border in column $i$ and $C_{\mathcal{B}}$ accepts at the bottom border in column $j < i$;
\item\label{enum:bottomafter} $C_{\mathcal{A}}$ accepts at the bottom border in column $i$ and $C_{\mathcal{B}}$ accepts at the bottom border in column $k \geq i$;
\end{enumerate}
\item\label{enum:rightright} $C_{\mathcal{A}}$ and $C_{\mathcal{B}}$ both accept at the right border.
\end{enumerate}
Depending on the guessed ``types" of the computations $C_{\mathcal{A}}$ and $C_{\mathcal{B}}$, the computation of $\mathcal{M}$ 
proceeds in one of the following ways:

\begin{figure}[t]
\centering
\begin{tikzpicture}[scale=0.55]
	\draw[draw=black] (0,0) rectangle (9,6);
	\draw[fill=black] (0.5,5.5) circle[radius=2pt];
	
	\draw[->, line width=1pt] (0.5,5.5) -- (1.5,5.5) -- (1.5,4.5) -- (3.5,4.5) -- (3.5,3.5) -- (5.5,3.5) -- (5.5,1) -- (6,1) -- (6,0);
	
	\draw[dashed, line width=0.5pt] (6,6) -- (6,0);
	
	\node at (7,0.75) {\footnotesize (1)};
	
	\node at (7,5.5) {\footnotesize col.\ $i$};
	
	\node at (3,2.5) {\footnotesize $C_{\mathcal{A}}$};
	\node at (4.5,-1) {Input word for $\mathcal{A}$};
\end{tikzpicture}
\hspace{1cm}
\begin{tikzpicture}[scale=0.55]
	\draw[draw=black] (0,0) rectangle (9,6);
	
	\draw[->, line width=1pt] (0.5,5.5) -- (1,5.5) -- (1,4.5) -- (2.5,4.5) -- (2.5,3) -- (3,3) -- (3,2.5) -- (6,2.5) -- (6,1.5) -- (8,1.5) -- (8,1) -- (9,1);
	
	\draw[dashed, line width=0.5pt] (6,6) -- (6,0);
	
	\draw[draw=black] (6,2.5) circle [radius=0.25];
	\node at (7, 3) {\footnotesize (2)};
	
	\node at (7,5.5) {\footnotesize col.\ $i$};
	
	\node at (2,1.5) {\footnotesize $C_{\mathcal{B}}$};
	\node at (4.5,-1) {Input word for $\mathcal{B}$};
\end{tikzpicture}
\caption{Illustration of Case 1 computation. The simulation of $C_{\mathcal{A}}$ accepts at (1). The computation of $\mathcal{M}$ will begin its second phase at (2) in the input word consisting of the row concatenation of input words for $\mathcal{A}$ and $\mathcal{B}$.}
\label{fig:case1illustrations}
\end{figure}
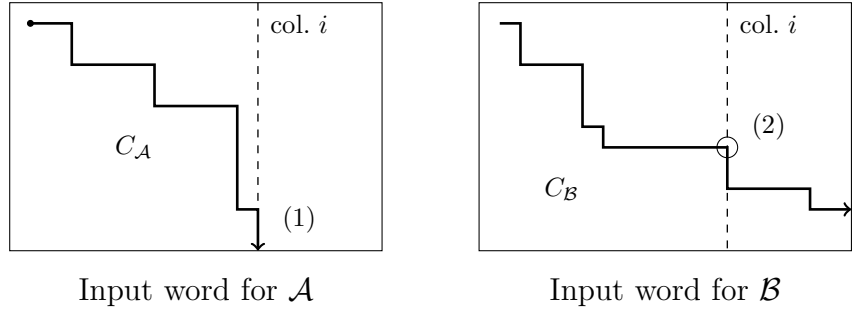

\paragraph{Case~\ref{enum:bottomright}.} The computation of $\mathcal{M}$ is divided into two phases. In the first phase, $\mathcal{M}$ simulates the computations of $\mathcal{A}$ and $\mathcal{B}$ in the following order:
\begin{enumerate}
\item[(i)] $\mathcal{M}$ simulates all possible downward moves of $\mathcal{A}$ by moving the input head and changing the state of $\mathcal{A}$;

\item[(ii)] $\mathcal{M}$ simulates all possible downward moves of $\mathcal{B}$ by moving the input head and changing the state of $\mathcal{B}$;

\item[(iii)] when $\mathcal{A}$ and $\mathcal{B}$ both make a rightward move, $\mathcal{M}$ simulates the move and changes the state of both $\mathcal{A}$ and $\mathcal{B}$.
\end{enumerate}
Note that, after completing steps (i) and (ii), at least one rightward move must occur in step (iii). After $\mathcal{M}$ completes step (iii), it continues from step (i).

When $\mathcal{M}$ simulates downward moves of the input head of $\mathcal{A}$ in step (i), it may nondeterministically guess that the input head of $\mathcal{A}$ has encountered a boundary symbol at the bottom border of the input word. If $\mathcal{A}$ is in an accepting state at that point, then $\mathcal{M}$ 
begins the second phase of its computation.

In the second phase, $\mathcal{M}$ simulates only the computation of $\mathcal{B}$. If $\mathcal{B}$ enters an accepting state when $\mathcal{M}$ encounters the right border, then $\mathcal{M}$ accepts.

Assume that $C_{\mathcal{A}}$ accepts at the bottom border in column $i$ of its input word. At the point when the first phase of the computation ends, the input head of $\mathcal{M}$ will be at the position corresponding to where $C_{\mathcal{B}}$ first enters column $i$. (See Figure~\ref{fig:case1illustrations}.) Although $\mathcal{M}$ performs its computation on the concatenated input, $C_{\mathcal{B}}$ performs its computation only on the input word to $\mathcal{B}$.

\paragraph{Case~\ref{enum:rightbottom}.} The first phase of the computation of $\mathcal{M}$ proceeds in the same manner as 
in the first phase of Case~\ref{enum:bottomright}. However, in this case, since there may be an unknown number of rows beneath the row in which $\mathcal{A}$ accepts, the input head of $\mathcal{M}$ need not be at the bottom border when $\mathcal{B}$ accepts at the bottom border.

Thus, during step (ii), $\mathcal{M}$ may nondeterministically guess that the input head of $\mathcal{B}$ has encountered a boundary symbol at the bottom border of the input word. 
Since $\mathcal{B}$ is IBR-accepting, we may assume $\mathcal{B}$ transitions immediately to its accepting state. At this point, $\mathcal{M}$ 
begins the second phase of its computation.

In the second phase, $\mathcal{M}$ simulates only the computation of $\mathcal{A}$. If $\mathcal{A}$ enters an accepting state when $\mathcal{M}$ encounters the right border, then $\mathcal{M}$ accepts.

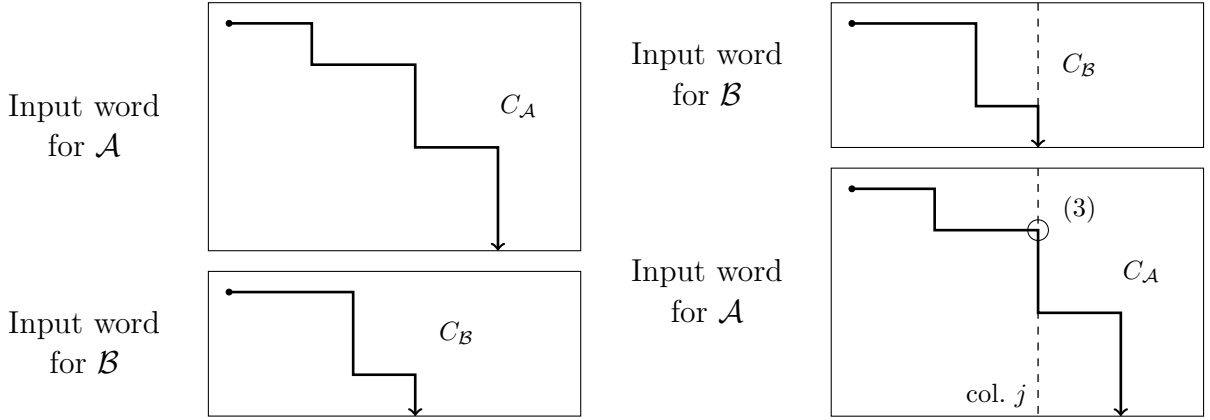
\begin{figure}[t]
\centering
\begin{tikzpicture}[scale=0.55]
	\draw[draw=black] (0,0) rectangle (9,6);
	\draw[fill=black] (0.5,5.5) circle[radius=2pt];
	
	\draw[->, line width=1pt] (0.5,5.5) -- (2.5,5.5) -- (2.5,4.5) -- (4.5,4.5) -- (5,4.5) -- (5,2.5) -- (7,2.5) -- (7,0);
	
	\node at (7.5,3.5) {\footnotesize $C_{\mathcal{A}}$};
	\node at (-3,3.45) {Input word};
	\node at (-3,2.55) {for $\mathcal{A}$};
	
	\draw[draw=black] (0,-0.5) rectangle (9,-4);
	\draw[fill=black] (0.5,-1) circle[radius=2pt];
	
	\draw[->, line width=1pt] (0.5,-1) -- (3.5,-1) -- (3.5, -2) -- (3.5, -3) -- (5,-3) -- (5,-4);
	
	\node at (6,-2) {\footnotesize $C_{\mathcal{B}}$};
	\node at (-3,-1.8) {Input word};
	\node at (-3,-2.7) {for $\mathcal{B}$};
\end{tikzpicture}
\hspace{0.25cm}
\begin{tikzpicture}[scale=0.55]
	\draw[draw=black] (0,0) rectangle (9,6);
	\draw[fill=black] (0.5,5.5) circle[radius=2pt];
	
	\draw[->, line width=1pt] (0.5,5.5) -- (2.5,5.5) -- (2.5,4.5) -- (4.5,4.5) -- (5,4.5) -- (5,2.5) -- (7,2.5) -- (7,0);
	
	\draw[dashed, line width=0.5pt] (5,6) -- (5,0);
	\draw[dashed, line width=0.5pt] (5,10) -- (5,6.5);
	
	\draw[draw=black] (5,4.5) circle [radius=0.25];
	\node at (6, 5) {\footnotesize (3)};
	
	\node at (4,0.5) {\footnotesize col.\ $j$};
	
	\node at (7.5,3.5) {\footnotesize $C_{\mathcal{A}}$};
	\node at (-3,3.45) {Input word};
	\node at (-3,2.55) {for $\mathcal{A}$};
	
	\draw[draw=black] (0,6.5) rectangle (9,10);
	\draw[fill=black] (0.5,9.5) circle[radius=2pt];
	
	\draw[->, line width=1pt] (0.5,9.5) -- (3.5, 9.5) -- (3.5, 8.5) -- (3.5, 7.5) -- (5,7.5) -- (5,6.5);
	
	\node at (6,8.5) {\footnotesize $C_{\mathcal{B}}$};
	\node at (-3,8.7) {Input word};
	\node at (-3,7.8) {for $\mathcal{B}$};
\end{tikzpicture}
\caption{Illustration of Case 3a computation. The left figure depicts the row concatenation, while the right figure depicts the ``swapped" computation. When the simulation of $C_{\mathcal{B}}$ accepts, the computation of $\mathcal{M}$ will begin its second phase at (3).}
\label{fig:case3aillustrations}
\end{figure}

\paragraph{Case~\ref{enum:bottombefore}.} 
We proceed in a similar manner as for Case~\ref{enum:bottomright}, but we modify the first phase of the computation of $\mathcal{M}$ by swapping steps (i) and (ii). 
Thus, in the first phase, $\mathcal{M}$ simulates the computations of $\mathcal{A}$ and $\mathcal{B}$ in the following order:
\begin{enumerate}
\item[(i$^{\prime}$)] $\mathcal{M}$ simulates all possible downward moves of $\mathcal{B}$ by moving the input head and changing the state of $\mathcal{B}$;

\item[(ii$^{\prime}$)] $\mathcal{M}$ simulates all possible downward moves of $\mathcal{A}$ by moving the input head and changing the state of $\mathcal{A}$;

\item[(iii$^{\prime}$)] when $\mathcal{A}$ and $\mathcal{B}$ both make a rightward move, $\mathcal{M}$ simulates the move and changes the state of both $\mathcal{A}$ and $\mathcal{B}$.\footnote{Note that step (iii$^{\prime}$) of Case~\ref{enum:bottombefore} is identical to step (iii) of Case~\ref{enum:bottomright}.}
\end{enumerate}

Since concatenation of unary words is a commutative operation, we may view this case as a simulation where the computation of $\mathcal{B}$ is performed before the computation of $\mathcal{A}$, allowing us to swap the first two steps of the first phase.

During step (i$^{\prime}$), $\mathcal{M}$ may nondeterministically guess that the input head of $\mathcal{B}$ has encountered a boundary symbol at the bottom border of the input word. Since $\mathcal{B}$ is IBR-accepting, we may assume $\mathcal{B}$ transitions immediately to its accepting state. At this point, $\mathcal{M}$ 
begins the second phase of its computation. (See Figure~\ref{fig:case3aillustrations}.)

In the second phase, $\mathcal{M}$ simulates only the computation of $\mathcal{A}$. If $\mathcal{A}$ enters an accepting state when $\mathcal{M}$ encounters the bottom border, then $\mathcal{M}$ accepts.

The logic behind determining the input head position after switching from simulating $C_{\mathcal{B}}$ to $C_{\mathcal{A}}$ is similar to the explanation given for Case~\ref{enum:bottomright}.

\paragraph{Case~\ref{enum:bottomafter}.} Analogous to the proof for Case~\ref{enum:bottomright}.

\paragraph{Case~\ref{enum:rightright}.} 
Since both $C_{\mathcal{A}}$ and $C_{\mathcal{B}}$ accept at the right border of their input words, there may be an unknown number of rows beneath the rows in which $\mathcal{A}$ and $\mathcal{B}$ accept. Therefore, the computation of $\mathcal{M}$ need only verify that its input word contains a sufficient number of rows to allow simulation of $C_{\mathcal{A}}$ and $C_{\mathcal{B}}$.

The computation of $\mathcal{M}$ proceeds in the same order as the steps outlined in the first phase of Case~\ref{enum:bottomright}. If, during step (iii), the input head of $\mathcal{M}$ encounters a boundary marker when both $\mathcal{A}$ and $\mathcal{B}$ are in accepting states, then $\mathcal{M}$ accepts.
\end{proof}
\end{theorem}

Closure under column concatenation follows by interchanging downward and rightward input head moves.

\begin{corollary}
Nondeterministic two-way two-dimensional automata over a unary alphabet are closed under column concatenation.
\end{corollary}


\section{Diagonal Concatenation}\label{sec:dconcatenation}

Anselmo et al.\ \cite{Anselmo2005NewOperations2DLanguages} introduced a new 
operation for unary two-dimensional words called ``diagonal concatenation". Given unary two-dimensional words $w$ and $v$ of dimension $m \times n$ and $m' \times n'$ respectively, the diagonal concatenation of those words, denoted $w \oslash v$, is a two-dimensional word of dimension $(m + m') \times (n + n')$ where $w$ is in the ``top-left corner" and $v$ is in the ``bottom-right corner".

\begin{figure}[t]
\[\arraycolsep=1.4pt\def\arraystretch{0.8}
w \oslash v = 
\begin{array}{cccccccc}
\#		& \#		& 		& \#			& \#		& 		& \# 		& \# \\
\#		& w_{1,1}	& \cdots	& w_{1,n} 	& x_{1,1}		& \cdots	& x_{1,n'}	& \# \\
		& \vdots	&		& \vdots	& \vdots		& 		& \vdots	& \\
\#		& w_{m,1}	& \cdots	& w_{m,n}	& x_{m,1}		& \cdots	& x_{m,n'}	& \# \\
\#		& y_{1,1}	& \cdots	& y_{1,n}	& v_{1,1}		& \cdots	& v_{1,n'}	& \# \\
		& \vdots	&		& \vdots	& \vdots		&		& \vdots	& \\
\#		& y_{m',1}	& \cdots	& y_{m',n}	& v_{m',1}		& \cdots	& v_{m',n'}& \# \\
\#		& \#		& 		& \#			& \#		& 		& \# 		& \#
\end{array}
\]
\caption{Diagonal concatenation of two-dimensional words}
\label{fig:2Ddiagonalconcatenation}
\end{figure}

In this section, we extend the diagonal concatenation operation to words over a general alphabet. In this case, the diagonal concatenation of two words $w$ and $v$, defined as before, produces a two-dimensional language consisting of words of dimension $(m + m') \times (n + n')$ where $w$ is in the top-left corner, $v$ is in the bottom-right corner, and words $x \in \Sigma^{m \times n'}$ and $y \in \Sigma^{m' \times n}$ are placed in the ``top-right" and ``bottom-left" corners of $w \oslash v$, respectively. 
The diagonal concatenation language is formed by adding to these corners all possible words $x$ and $y$ over $\Sigma$. 
An example word from such a language is depicted in Figure~\ref{fig:2Ddiagonalconcatenation}. 
We may define the diagonal concatenation of two languages $A$ and $B$ in a similar manner as for row and column concatenation: 
the top-left corner contains only words from $A$, and the bottom-right corner contains only words from $B$.

\begin{remark}
Note that an automaton recognizing $w \oslash v$ only needs to read the contents of the top-left and bottom-right corners to determine whether a word is in the diagonal concatenation language. Thus, we may add any symbols from $\Sigma$ to the top-right and bottom-left corners to ensure the resulting word is a contiguous rectangle. If an automaton recognizes the diagonal concatenation language as defined earlier, where all possible words $x$ and $y$ are placed in the top-right and bottom-left corners, respectively, then it will recognize any word with $w$ and $v$ in the appropriate locations, as desired.
\end{remark}

Before we present the main results, we will prove a small result about diagonal concatenation where individual words are separated by additional boundary markers. As one might expect, if an automaton is able to determine where one word ends and another word begins, then closure follows easily.

\begin{theorem}
All two-dimensional automaton models are closed under diagonal concatenation where words are separated by boundary markers.

\begin{proof}
Suppose we are given a pair of two-dimensional automata $\mathcal{A}$ and $\mathcal{B}$ recognizing languages $A$ and $B$, respectively. We may construct a new automaton $\mathcal{C}$ to recognize the diagonal concatenation language $A \oslash B$ in the following way.

Convert $\mathcal{A}$ to an IBR-accepting automaton, denoted $\mathcal{A}'$. Then, simulate the computation of $\mathcal{A}'$ with $\mathcal{C}$. If $\mathcal{A}'$ accepts at the bottom border of the input word, then the input head of $\mathcal{C}$ moves downward once and moves rightward until it crosses a boundary symbol and encounters an alphabet symbol. Otherwise, if $\mathcal{A}'$ accepts at the right border, then the input head of $\mathcal{C}$ moves rightward once and moves downward until it crosses a boundary symbol and encounters an alphabet symbol. In either case, $\mathcal{C}$ proceeds to simulate the computation of $\mathcal{B}$, and in all cases, $\mathcal{C}$ can be made to be of the same type as both $\mathcal{A}$ and $\mathcal{B}$.
\end{proof}
\end{theorem}


\subsection{Two-Way Two-Dimensional Automata}\label{subsec:dconcatenationtwoway}

For the two-way two-dimensional automaton model, the input head  
is able to recognize when it reaches the bottom or right border of its input word. However, since the input head cannot move upward or leftward, it cannot leave the border once it moves onto a boundary symbol. If the input head makes further moves upon reaching the border, it can only read boundary symbols until the automaton halts.

Anselmo et al.\ \cite{Anselmo2005NewOperations2DLanguages} state that their definition of a two-way two-dimensional automaton is equivalent to a two-tape one-dimensional automaton whose input heads 
only move rightward. This suggests that their model can detect when it has reached not only one, but both of its input word's bottom and right borders, giving it 
more recognition power than our model, which is only able to determine when it has reached either the bottom or right border of its input word, but not both simultaneously. In terms of boundary symbols, the model of Anselmo et al.\ is 
akin to placing a distinguished boundary symbol at the bottom-right corner of the border of the input word.

Using our two-way model, where all boundary symbols are identical, we obtain the following closure result for diagonal concatenation.

\begin{theorem}\label{thm:2Wdiagonalconcatenationclosure}
Nondeterministic two-way two-dimensional automata over a general alphabet are closed under diagonal concatenation.

\begin{proof}
Suppose we are given two nondeterministic two-way two-dimensional automata $\mathcal{A}$ and $\mathcal{B}$ recognizing languages $A$ and $B$, respectively. We may construct a new automaton $\mathcal{C}$ to recognize the language $A \oslash B$ in the following way.

Begin by converting $\mathcal{A}$ to an IBR-accepting automaton, denoted $\mathcal{A}'$. Then, simulate the computation of $\mathcal{A}'$ with $\mathcal{C}$, but modify the transition function so that the simulation performed by $\mathcal{C}$ accepts if and only if $\mathcal{A}'$ would accept upon reading a boundary marker. In this way, $\mathcal{C}$ is pretending to read a boundary marker that would surround a word from $A$, but that does not appear within words from $A \oslash B$.

At this stage in the computation, the input head of $\mathcal{C}$ will have made either a downward move or a rightward move, depending on whether $\mathcal{A}'$ accepts its input word at the bottom border or the right border, respectively. If the input head of $\mathcal{C}$ previously made a downward move, then the input head will move rightward some nondeterministically-selected number of symbols. If the input head of $\mathcal{C}$ previously made a rightward move, then the input head will move downward in a similar fashion. In either case, $\mathcal{C}$ begins to simulate the computation of $\mathcal{B}$ after these nondeterministic moves.
\end{proof}
\end{theorem}

Evidently, the power of nondeterminism is crucial for the automaton $\mathcal{C}$ to recognize diagonally-concatenated words. Indeed, if we remove nondeterminism, then we also lose closure.

\begin{theorem}\label{thm:2Wdeterministicdiagonalconcatenationclosure}
Deterministic two-way two-dimensional automata over a general alphabet are not closed under diagonal concatenation.

\begin{proof}
Define a language $L$ as the set of two-dimensional words over the alphabet $\Sigma = \{\texttt{0}, \texttt{1}\}$ where the top-left symbol of each word is \texttt{1}. A two-dimensional automaton can recognize this language by immediately reading the symbol at the initial position of its input head.

Suppose there exists a deterministic two-way two-dimensional automaton $\mathcal{A}$ that recognizes the language $L \oslash L$. Any accepting computation of $\mathcal{A}$ must visit at least two occurrences of the symbol \texttt{1}, corresponding to the top-left symbols in both words of the concatenation.

Consider the computation of $\mathcal{A}$ on an input word $w$ of dimension $3 \times 3$, where the top-left symbol of $w$ is \texttt{1} and all other symbols are \texttt{0}. Clearly, $w \not\in L \oslash L$, so the computation of $\mathcal{A}$ will reject. However, since $\mathcal{A}$ is a two-way two-dimensional automaton, the input head cannot visit all symbols in the bottom-right $2 \times 2$ subword of $w$. Thus, we may change an unvisited symbol from \texttt{0} to \texttt{1} to obtain a word $w'$ that belongs to $L \oslash L$, but is not accepted by $\mathcal{A}$.
\end{proof}
\end{theorem}


\subsection{Three-Way Two-Dimensional Automata}\label{subsec:dconcatenationthreeway}

Given the result of Theorem~\ref{thm:2Wdeterministicdiagonalconcatenationclosure}, we should expect not to obtain a positive closure result for deterministic three-way two-dimensional automata. However, 
unlike Theorem~\ref{thm:2Wdeterministicdiagonalconcatenationclosure}, we are considering three directions of movement, and so 
we cannot assert that the input head of 
our automaton is incapable of reading all symbols within 
the input word. (Indeed, the input head 
of a three-way two-dimensional automaton 
may read all symbols via a left-to-right sweeping motion.) Thus, we require a different approach.

Under certain conditions, a two-way one-dimensional automaton $\mathcal{N}$ can simulate the computation of a three-way two-dimensional automaton $\mathcal{M}$ on a particular row of its input word. Moreover, 
as we will see, 
the number of states of $\mathcal{N}$ depends linearly on the number of states of $\mathcal{M}$.

Here, we construct a diagonal concatenation language $A \oslash B$, where $A$ and $B$ are languages recognized by deterministic three-way two-dimensional automata $\mathcal{A}$ and $\mathcal{B}$, and then we show that there exists no two-way one-dimensional automaton $\mathcal{C}$ with enough states to simulate the computation of $\mathcal{A}$ and $\mathcal{B}$ together on the diagonal concatenation language.

To arrive at the main result, we first require two technical lemmas.

\begin{lemma}\label{lem:distancefromboundary}
Let $\mathcal{M}$ be a deterministic three-way two-dimensional automaton with $n$ states. Consider the computation of $\mathcal{M}$ on an input word over the alphabet $\Sigma = \{\texttt{0}, \texttt{1}\}$, where row $i$ consists entirely of \texttt{0}s.

If the input head of $\mathcal{M}$ visits the first or last symbol of row $i$ and moves downward to row $i+1$, then this move happens at distance at most $n+1$ from one of the boundary markers.

\begin{proof}
If $\mathcal{M}$ enters row $i$ in the first or last column and moves downward immediately after, then the result follows. If $\mathcal{M}$ enters row $i$, visits the first or last symbol of the row, and makes fewer than $n$ leftward/rightward moves before moving downward, then the result also follows.

Since $\mathcal{M}$ must visit either the first or last symbol of row $i$, it can make at most $n$ moves leftward/rightward without entering a loop. If $\mathcal{M}$ does not move downward to row $i+1$ within the first $n$ leftward/rightward moves, then it will be forced to move leftward/rightward in a loop until it reaches the other end of row $i$.

Therefore, a downward move may only occur within distance $n$ from the first or last symbols of the row, and thus may only occur within distance $n+1$ from one of the boundary markers.
\end{proof}
\end{lemma}

Given a deterministic three-way two-dimensional automaton $\mathcal{M}$, consider the computation of $\mathcal{M}$ on row $i$ of its input word. We say that a two-way one-dimensional automaton $\mathcal{N}$ correctly simulates the computation of $\mathcal{M}$ on row $i$ if, given row $i$ as input, $\mathcal{N}$ accepts if and only if $\mathcal{M}$ moves downward from row $i$.

\begin{lemma}\label{lem:1Dsimulatedownward}
Suppose that a deterministic three-way two-dimensional automaton $\mathcal{M}$ has $n$ states, and that $\mathcal{M}$ enters row $i$ of its input word at distance at most $n+1$ from one of the boundary markers. Then there exists a deterministic two-way one-dimensional automaton $\mathcal{N}$ with at most $2n+3$ states that correctly simulates the computation of $\mathcal{M}$ on row $i$.

\begin{proof}
By our assumption, $\mathcal{M}$ begins its computation at distance at most $n+1$ from one of the boundary markers of row $i$. On the other hand, $\mathcal{N}$ begins its computation at the leftmost position of its input.

We require at most $n+2$ states to move the input head of $\mathcal{N}$ to the initial position of the input head of $\mathcal{M}$ after it enters row $i$; $n+1$ states are used to count the number of moves made by $\mathcal{N}$, and one state is required to move the input head of $\mathcal{N}$ rightward if $\mathcal{M}$ entered row $i$ at one of the rightmost $n+1$ positions. At this point, $\mathcal{N}$ directly simulates the computation of $\mathcal{M}$ using $n$ states. Once $\mathcal{M}$ follows a transition with a downward move, $\mathcal{N}$ enters a designated accepting state. Altogether, this construction for $\mathcal{N}$ requires at most $2n+3$ states.
\end{proof}
\end{lemma}

Kapoutsis \cite{Kapoutsis2005RemovingBidirectionality} proved that, given a deterministic two-way one-dimensional automaton with $n$ states, we may convert it to an equivalent deterministic one-way one-dimensional automaton with $h(n) = n(n^{n} - (n-1)^{n})$ states. We will use this value $h(n)$ in the proof of the main result: given languages $A$ and $B$ that are recognized by deterministic three-way two-dimensional automata, the language $A \oslash B$ need not be recognized by the same model.

\begin{theorem}\label{thm:3Wdeterministicdiagonalconcatenationclosure}
Deterministic three-way two-dimensional automata over a general alphabet are not closed under diagonal concatenation.

\begin{proof}
Let $\Sigma = \{\texttt{0}, \texttt{1}\}$. Let $A$ be the language of $1 \times n'$ two-dimensional words where $n' \geq 1$ and the single row consists entirely of \texttt{0}s. Let $B$ be the language of $3 \times n''$ two-dimensional words where $n'' \geq 1$ and all three rows consist entirely of \texttt{0}s, apart from the top-left corner symbol of each word, which is \texttt{1}.

Suppose there exists a deterministic three-way two-dimensional automaton $\mathcal{C}$ with $n$ states recognizing the language $A \oslash B$. Each word in $A \oslash B$ consists of exactly four rows. Since $\mathcal{C}$ can remember the number of rows it visited, we may assume without loss of generality that, 
when $\mathcal{C}$ moves downward from row 4, it accepts. Any moves to a rejecting state 
are simulated by ``stay-in-place" moves.

Let $k = h(2n+3) + 1$. Consider the set of two-dimensional words $X$ where each word has dimension $4 \times 2k$, the first row of each word consists entirely of \texttt{0}s, the second row is of the form $\texttt{0}^{k}\texttt{1}\texttt{0}^{k-1}$, the third row consists entirely of \texttt{0}s, and the fourth row is of the form $\{\texttt{0}, \texttt{1}\}^{2k}$. Evidently, given a word $w \in X$, we also have that $w \in A \oslash B$ if and only if the last $k$ symbols of row 4 are all \texttt{0}.

Consider any accepting computation of $\mathcal{C}$ on a word $w \in X$. During this computation, $\mathcal{C}$ must visit the last symbol of the third row. Otherwise, we could change the last symbol of the third row to \texttt{1}, and $\mathcal{C}$ would accept a word not belonging to $A \oslash B$.

Since $\mathcal{C}$ visits the last symbol of the third row, by Lemma~\ref{lem:distancefromboundary}, we know that the input head must subsequently enter the fourth row at distance at most $n+1$ from the rightmost boundary marker. Then, by Lemma~\ref{lem:1Dsimulatedownward}, there exists a deterministic two-way one-dimensional automaton $\mathcal{D}$ with $2n+3$ states that correctly simulates the computation of $\mathcal{C}$ on the fourth row. We may convert $\mathcal{D}$ to an equivalent one-way one-dimensional automaton $\mathcal{D}'$ with $h(2n+3)$ states. From here, $\mathcal{D}'$ must check that each of the rightmost $k = h(2n+3)+1$ symbols in the fourth row is \texttt{0}. 
But, since $\mathcal{D}'$ has $k - 1 = h(2n+3)$ states and the fourth row consists of $2k$ symbols, $\mathcal{D}'$ is unable to count up to the $k$th symbol in order to determine where the latter $k$ symbols in that row begin.
\end{proof}
\end{theorem}


\section{Conclusion}

In this paper, we considered closure properties of various concatenation operations on two-dimensional automata. We showed that 
two-way two-dimensional automata over a general alphabet are not closed under either row or column concatenation. 
For a unary alphabet, on the other hand, we showed that nondeterministic two-way two-dimensional automata are closed under both row and column concatenation. 
We further 
showed that nondeterministic two-way two-dimensional automata over a general alphabet are closed under diagonal concatenation, while neither 
deterministic 
two-way nor deterministic three-way two-dimensional automata are closed.

There remain some open problems related to two-dimensional concatenation. Most closure results for concatenation assume the use of a general alphabet. Studying concatenation for unary two-dimensional automaton models in particular could prove interesting. Furthermore, 
nothing is yet known about the closure of diagonal concatenation on four-way two-dimensional automata. 
Lastly, we conjecture that nondeterministic three-way two-dimensional automata over a general alphabet are not closed under diagonal concatenation. However, showing this would require essentially a different proof than that given for Theorem~\ref{thm:3Wdeterministicdiagonalconcatenationclosure}.


\bibliographystyle{plain}
\bibliography{./References.bib}


\end{document}